\documentclass[12pt]{iopart}
\usepackage[dvips]{graphicx}

\newcommand{\la}[1]{\label{#1}}
\newcommand{\be}{\begin{equation}}
\newcommand{\ee}{\end{equation}}
\newcommand{\ba}{\begin{eqnarray}}
\newcommand{\ea}{\end{eqnarray}}
\newcommand{\fig}{Fig.~}

\newcommand{\nn}{\nonumber \\}

\newcommand{\eq}[1]{Eq.~(\ref{#1})}
\makeatletter
\renewcommand\section{\@startsection {section}{1}{\z@}%
                                   {-5.5ex \@plus -1ex \@minus -.2ex}
                                   {2.3ex \@plus.2ex}%
                                   {\normalfont\large\bfseries}}
\renewcommand\subsection{\@startsection{subsection}{2}{\z@}%
                                     {-3.25ex\@plus -1ex \@minus -.2ex}%
                                     {1.5ex \@plus .2ex}%
                                     {\normalfont\normalsize\bfseries}}
\renewcommand\thesection {\@arabic\c@section}
\renewcommand\thesubsection   {\thesection.\@arabic\c@subsection}
\renewcommand{\@seccntformat}[1]{%
\csname the#1\endcsname.\hspace{1.0em}}
\makeatother

\begin{document}


\title{Scalar condensate decay in a fermionic heat bath in the early
    universe}

\author{Kari Enqvist and Janne H\"ogdahl}

\address{Department of Physical Sciences, University of Helsinki \\ and
Helsinki Institute of Physics, \\ P. O. Box 64, FIN-00014 University of
Helsinki, Finland.}

\begin{abstract}
We consider one-loop thermal effects on the decay of a scalar
field zero mode initially dominating the energy density of the
universe. We assume fermionic decay channels and take into account
the effects due to both particle and hole excitations, and present
approximate expressions for the absorption and decay rates. We
apply the results to the inflaton and solve the Boltzmann
equations to find the temperature evolution of the fermionic
plasma. We show that the reheat temperature can be greater than
the inflaton mass if the inflaton decays into more than one
fermionic species.
\end{abstract}

\eads{\mailto{kari.enqvist@helsinki.fi}, \mailto{janne.hogdahl@helsinki.fi}}

\maketitle

%
\section{Introduction}

Scalar condensates often play an important role in cosmology. For
example, the superluminal expansion of the early inflationary
universe is driven by primordial dark energy, which in scalar
field inflation is the energy of the zero mode inflaton condensate
\cite{Lyth:1998xn}. At the end of inflation the condensate $\phi$
oscillates and then decays roughly speaking when the decay rate
$\Gamma_\phi > H$, where $H$ is the Hubble rate. The decay
products are assumed to thermalize almost instantaneously by
scattering against each other, thus reheating the cold universe.
While the inflaton condensate is decaying, the expansion rate of
the universe depends on the varying relative magnitudes of the
plasma and the condensate energies, which thus affect the decay
and reheating processes. It turns out
\cite{Scherrer:1984fd,Chung:1998rq,Giudice:2000ex} that there is a
maximum temperature $T_{MAX}$ which the plasma can attain; this is
reached when the plasma energy is still subdominant.  The reheat
temperature $T_{RH}$ marks the time after which the universe is
dominated by the relativistic gas of the decay products.

The process of thermalization of a scalar condensate in the heat
bath of its decay products in an expanding universe is a generic
problem that is relevant not only for inflation but, for example,
also for the decay of the squark/slepton condensate of the
minimally supersymmetric standard model (MSSM)
\cite{Enqvist:2003gh}. In thermalized plasma the dispersion
relations of the decay products are modified by what is commonly
known as plasma masses \cite{Weldon:bn} (the role of plasma masses
in early universe was first noted in \cite{Linde:gh}). These can
affect the decay rate, as was recently discussed in
\cite{Kolb:2003ke} in the context of inflation. There the plasma
masses were put in by hand into the decay rate $\Gamma_\phi$ to
provide a kinematic suppression factor. The authors found that the
inclusion of the plasma masses gives rise to an era of constant
temperature during reheating; this happens because the masses of
the decay products of the inflaton increase with the temperature
and at some point the energy of the decay products would exceed
the energy of the inflaton zero mode, rendering the decay
kinematically forbidden.

However, a thermal background induces modifications in the scalar condensate
decay rate which are more subtle than a simple phase space suppression. If one
assumes that the scalar condensate decays mainly into fermions, as we will do
in this paper, the plasma has two kinds of elementary fermionic excitations,
called particles and holes. Particles are modes which in the limit $T\to 0$
yield ordinary elementary excitations, whereas holes are collective modes
present in thermal bath only; they are called "holes" since these excitations
correspond to the removal of an antiparticle, creating a state with all the
same quantum numbers as a particle state, but with different energy
\cite{Weldon:1989ys}.  Thus for instance an inflaton condensate can decay into
a pair of particles or a pair of holes, or it can transform into a particle by
absorbing a hole from the thermal bath.  All these processes contribute to the
total thermalization rate of the inflaton condensate and need to be evaluated.

In this paper we calculate the thermalization rate of a scalar
condensate decaying into fermionic particles at finite
temperature. The results could be applied to any situation where
there is an initially dominant scalar field decaying in an
expanding universe. Here we focus on the reheating phase of the
early universe.

%
\section{Finite temperature thermalization rates}

Let us consider the interaction Lagrangian
\begin{equation}
{\cal L}= g_{Y}\phi\bar\psi\psi~,
\end{equation}
where $\phi$ is a some scalar field that forms a zero mode condensate in the
early universe, and $\psi$ is a fermion. Here $g_{Y}$ is the Yukawa coupling
between the scalar and the fermion. In what follows we will neglect the
self-couplings of $\phi$ as well as its couplings to other scalars and assume
that the fermions $\psi$ are brought to thermal equilibrium instantaneously
after they are produced by virtue of $\psi\psi$ scatterings. We also assume
that the rest masses of the fermions are much smaller than the ambient
temperature $T$, so we can take the fermionic dispersion relation to be
dominated by the first order temperature correction $m=CgT$, where $g$ is a
coupling constant appropriate for $\psi\psi$ scattering and $C$ is some
number, which we take to be 1 for simplicity.

The thermalization rate $\Gamma_\phi(T)$ of a scalar condensate
has been computed in the past (see e.g. \cite{Elmfors:re}) and is
given by the imaginary part of the zero mode propagator $\Sigma$;
at one-loop level this is the cut of the relevant one-loop
diagrams at finite $T$. In general one can write
\begin{equation}
\Gamma_\phi(T,\omega) = - \frac{\textrm{Im} {\Sigma}(\omega,
T)}{\omega}, \label{gammat}
\end{equation}
where $\omega$ is the energy of the given mode. $\Sigma$ receives
contributions from all the fermionic modes present at finite temperature.

When considering the thermalization of a scalar field, there are
two kinds of processes. First, the scalar zero mode can decay
directly into a pair of particles or a pair of holes. The energies
of the produced particles are then constrained by
\begin{equation}
M_{\phi} = 2 \omega_{i}(k)~;~i=p,h~,
\end{equation}
where $p$ denotes particle and $h$ hole. Mostly we will be interested in a
cold condensate dominating the universe, such as the inflaton or the MSSM flat
direction condensate. Hence we may ignore thermal corrections to the scalar
mass and take $\omega=M_{\phi}$.

Second, if the mass of the scalar is small enough (or temperature
large enough), the possibility opens for a hole in the heat bath
to absorb the scalar to produce a particle. This could be called
thermal fragmentation of the condensate. In this case the energies
are constrained by
\begin{equation}
M_{\phi} + \omega_{h}(k) = \omega_{p}(k).
\end{equation}
As will be discussed below, it turns out that for thermal fragmentation to
occur, the temperature of the thermal bath has to become larger than the
scalar rest mass. However, since the scalar-fermion (heat bath) coupling,
$g_{Y}$, is typically much smaller than the coupling $g$ relevant for
$\psi\psi$ scattering, we can ignore the thermal mass of the scalar.

The dispersion relations for fermionic positive energy hole and particle
excitations with thermal masses $m=gT$ can be written as
\cite{Weldon:1989ys,Elmfors:re}
\ba
\hat{\omega}_{p}-\hat{k}-\frac{g^2}{\hat{k}}-\frac{g^2}{2\hat{k}} \left(
1-\frac{\hat{\omega}_{p}}{\hat{k}} \right) \ln \Bigl| \frac{\hat{\omega}_{p}
+\hat{k}}{\hat{\omega}_{p}-\hat{k}} \Bigr| = 0 \label{partdisp}~,\\
\hat{\omega}_{h}+\hat{k}+\frac{g^2}{\hat{k}}-\frac{g^2}{2\hat{k}}
\left( 1+\frac{\hat{\omega}_{h}}{\hat{k}} \right) \ln \Bigl|
\frac{\hat{\omega}_{h} +\hat{k}}{\hat{\omega}_{h}-\hat{k}} \Bigr|
= 0~,\label{holedisp}
\ea
where $\hat{\omega}=\omega/T$, $\hat{k}=k/T$. The dispersion relation for
holes can be also expressed as \cite{Weldon:1989ys}
\begin{equation}
\hat{\omega_{h}} = \hat{k}
\coth \left[ \frac{\hat{k}^{2}}{g^{2}} + \frac
{\hat{k}}{\hat{\omega}_{h} + \hat{k}} \right];\label{toinenmuoto}
\end{equation}
this form turns out to be more convenient in numerical calculations. The
relations Eqs.  (\ref{partdisp})-(\ref{toinenmuoto}), together with the energy
conditions for absorption and decay, can be solved numerically to obtain
$\omega_h$, $\omega_p$ and $k$ as functions of temperature, and the results
can be used to find $\Gamma_\phi(T)$. The imaginary parts needed for the decay
rates were computed in \cite{Elmfors:re} and read ($t=T/M_{\phi}$)
\begin{equation}
\textrm{Im} \Sigma_{\textrm{abs}} = - \frac{g_{Y}^{2}}{2 \pi g^{4}} \left[
\hat{k}^{2} t^{2} (\hat{\omega}_{p}^{2} - \hat{k}^{2}) (\hat{\omega}_{h}-
\hat{k}^{2}) \left( n_{h}-n_{p} \right) \right] M_{\phi}
\end{equation}
for the case of absorption, whereas
\begin{equation}
\fl \textrm{Im}
\Sigma_{\textrm{dec}} = - \frac{g_{Y}^{2}}{4 \pi g^{4}} \left[
\hat{k}^{2} t^{2} (\hat{\omega}_{p}^{2}-\hat{k}^{2})^{2} \left( 1-
2 n_{p} \right)+\hat{k}^{2} t^{2}
(\hat{\omega}_{h}^{2}-\hat{k}^{2})^{2} \left( 1- 2 n_{h} \right)
\right] M_{\phi}
\end{equation}
for the case of decay; here $g_{Y}$ is the Yukawa coupling between the scalar
and the fermion and
\begin{equation}
n_{h,p} = \frac{1}{\exp (\hat{\omega}_{h,p}) +1}
\end{equation}
is the distribution function.

The total scalar condensate thermalization rate
$\Gamma_\phi(T,\omega) \equiv \Gamma_{abs}+ \Gamma_{dec}$ is a
complicated function depending on the coupling constant $g$.
However, varying $g$ and fitting the numerical results with simple
functional forms we find that the thermalization rates are well
described by
\ba \Gamma_{dec} (t,g) &=& \frac{1}{4 \pi}(4-6gt)
\left( a e^{-b(gt)^2}+
c \right) \left( 1- \frac{2}{e^{1/(2t)}+1} \right)~, \\
\Gamma_{abs} (t,g) &=& \left( (D_2g^2+D_1g+D_0)t + \frac{E}{g}
\right)^{-1}~, \ea
where $a=0.34;~b=5.6546;~c=-0.09$ and
$D_2=3.746;~D_1=-0.7058;~D_0=11.1278;~E=1.8015$.

Since the direct decay channel is the only possible one when $T$ is small
enough, while the absorption channel opens only when $T$ is sufficiently
large, there is a range of temperatures where the both reactions are
kinematically forbidden if the scalar couples only to a single fermion.  In
reality the condensate is more than likely to couple to many species of
fermions. For instance, the MSSM flat direction condensate typically couples
to meny quark and/or lepton generations. Likewise, the inflaton could well
couple to several fermionic species. If a condensate couples to more than one
species with different thermal masses, it is possible to obtain a piecewise
continuous non-zero decay rate. This is because the end point of the direct
decay channel and the starting point of the absorption channel depend on the
coupling constant: if the scalar condensate decays directly to a fermion with
a thermal mass proportional to a coupling constant $g_{1}$ and absorbs a
fermion with a thermal mass proportional to a coupling constant $g_{2}$, then
in order for the temperature gap in the thermalization rate to vanish the
ratio of couplings must satisfy the relation
\begin{equation}
{\frac{g_{2}}{g_{1}}} < 0.25~,
\end{equation}
which is easily realized in actual models.  The Yukawa couplings may of course
also vary, but here we keep all the Yukawas equal for simplicity.

To demonstrate the interplay of the decay and absorption channels in case of
more than one fermion species, we have calculated the decay rates for a scalar
coupled to two fermions with thermal masses proportional to coupling constants
$g_{1}=1$, $g_{2}=0.1$ and $g_{1}=1$, $g_{2}=0.2$; the results are plotted in
\fig\ref{fig:gammas}.

\begin{figure}[t]

\begin{center}
\includegraphics[width=5.5cm,angle=270]{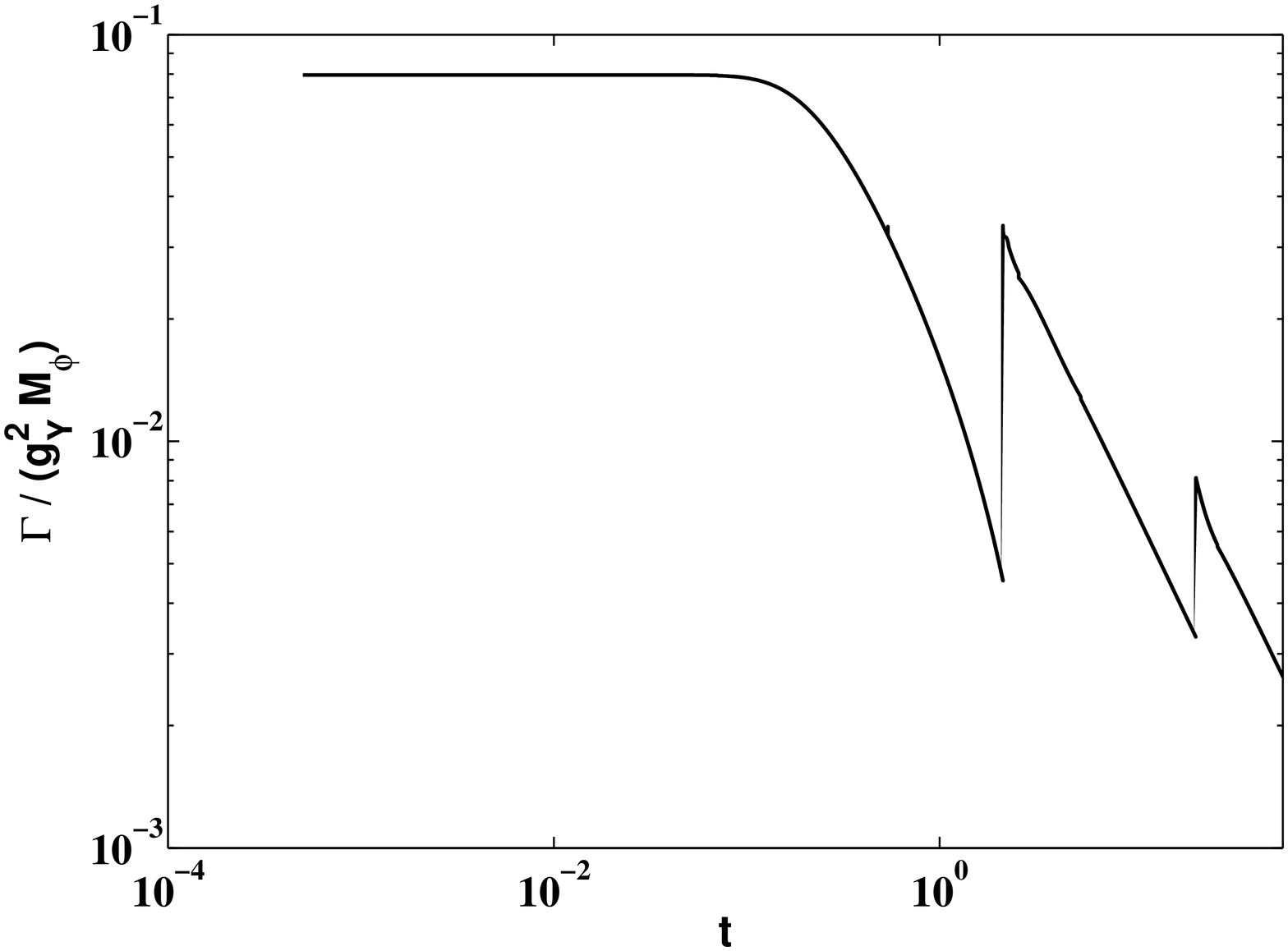}%
\hspace*{0.5cm}%
\includegraphics[width=5.5cm,angle=270]{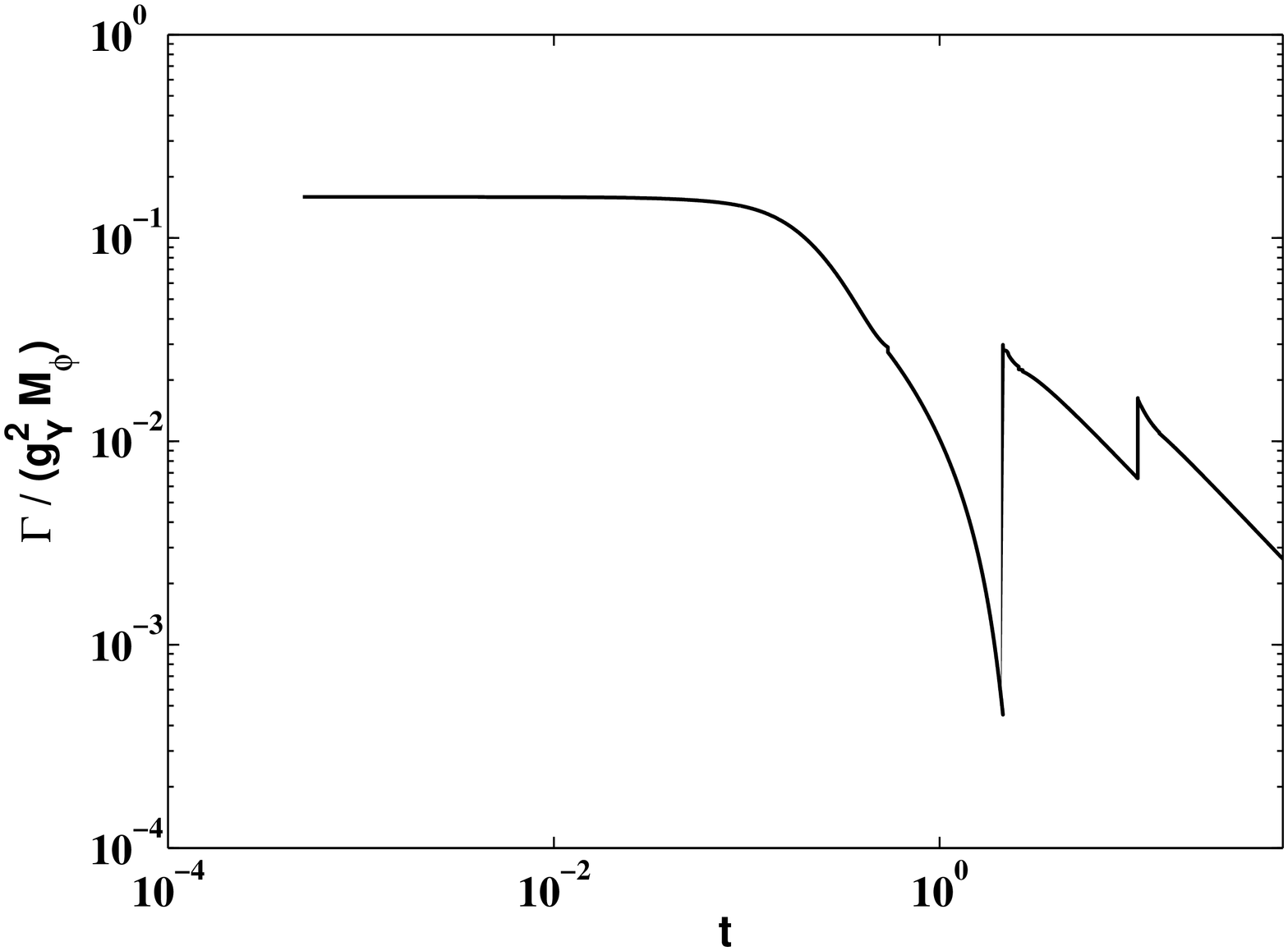}%
\end{center}

\caption[a]{The complete thermalization rates for a scalar zero mode decaying
into two fermionic particles with plasma masses proportional to $g_{1}=1$ and
$g_{2}=0.1$ (left), and $g_{1}=1$ and $g_{2}=0.2$ (right). The low temperature
part of the plot corresponds to the direct decay channels, while the zigzag
patterns are due to absorption channels opening first to particles with the
largest thermal mass, then to particles with a smaller thermal mass.}

\la{fig:gammas}
\end{figure}


\section{Solving the Boltzmann equations}

The energy partitioning between the scalar condensate and the
relativistic heat bath and its time evolution in the expanding
universe are described by the Boltzmann equations,
\ba
\dot{\rho}_{\phi}+3H \rho_{\phi}+\Gamma_{\phi}(T) \rho_{\phi} =0~,
\nn \dot{\rho}_{R}+4H \rho_{R}-\Gamma_{\phi}(T) \rho_{\phi} =0~,
\label{boltz} \ea
where $\rho_{\phi}$ and ${\rho}_{R}$ are
respectively the inflaton and the radiation energy densities.
Using the dimensionless quantities
\begin{equation}
\Phi \equiv \rho_{\phi}
M_{\phi}^{-1} a^{3} \, ; \,\, R \equiv \rho_{R} a^{4}
\end{equation}
as was done in e.g. \cite{Kolb:2003ke}, and defining a new variable $x=a
M_{\phi}$, \eq{boltz} can be written in a form more suitable for numerical
calculations ($' \equiv d/dx$),
\ba
\Phi' & =
&-\sqrt{\frac{3}{8\pi}}\frac{M_{Pl}}{M_{\phi}^{2}}
\Gamma_{\phi}(T)
\frac{x}{\sqrt{\Phi x+R}} \Phi~,\\
R' & = & \sqrt{\frac{3}{8\pi}}\frac{M_{Pl}}{M_{\phi}^{2}}
\Gamma_{\phi}(T) \frac{x^{2}}{\sqrt{\Phi x+R}} \Phi~.
\ea
Assuming that at the beginning the condensate energy density dominates, the
initial conditions at $x=x_{I}$ are given by $R(x_{I})=0$ and $\Phi
(x_{I})=\Phi_{I}$, where $\Phi_{I}$ is obtained from the Friedmann equation,
\begin{equation}
\Phi_{I}=\frac{3}{8 \pi}
\frac{M_{Pl}^{2}}{M_{\phi}^{2}} \frac{H_{I}^{2}}{M_{\phi}^{2}}
x_{I}^{3}~.
\end{equation}
\begin{figure}[t]

\begin{center}
\includegraphics[width=5.5cm,angle=270]{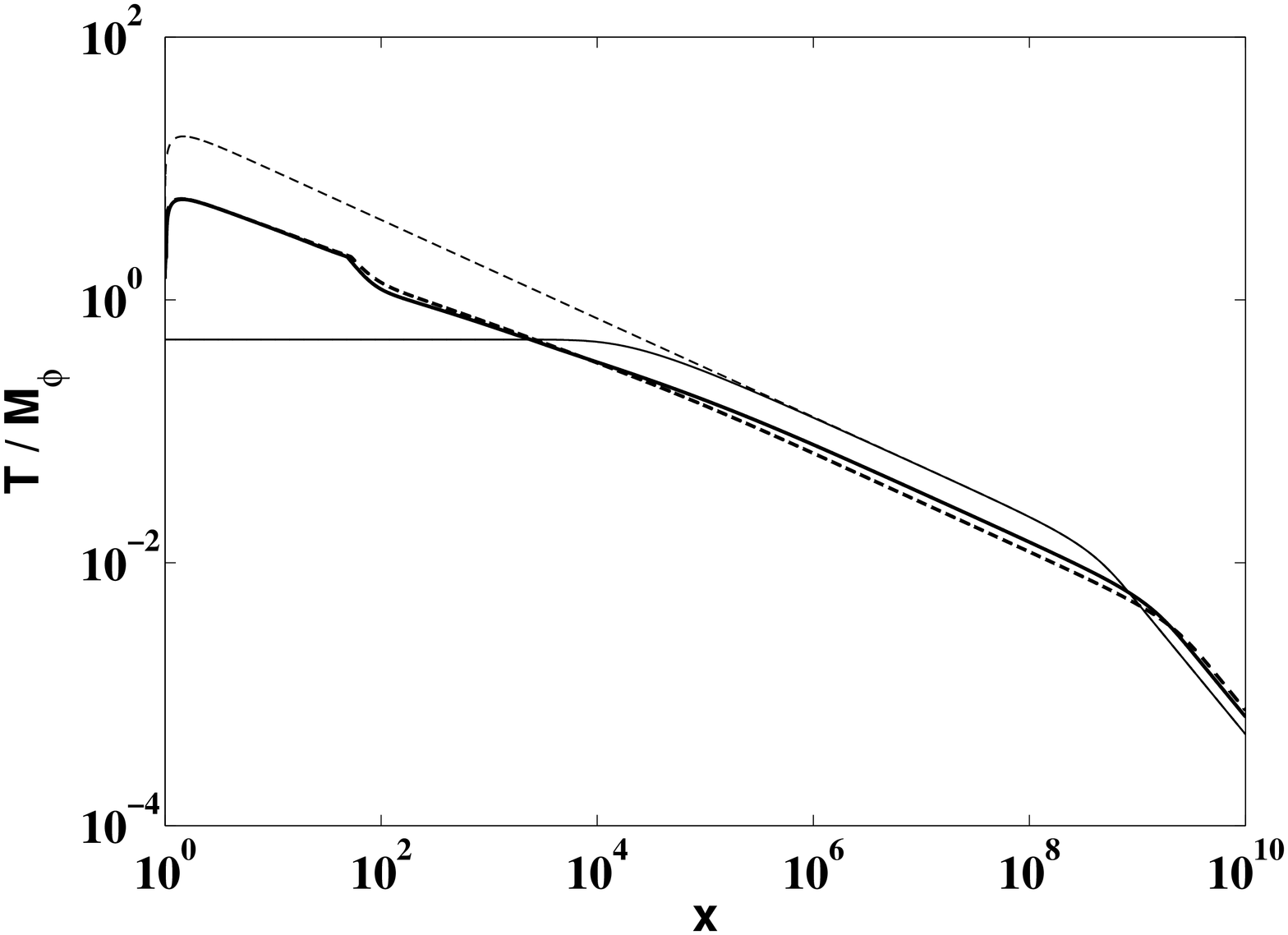}%
\hspace*{0.5cm}%
\includegraphics[width=5.5cm,angle=270]{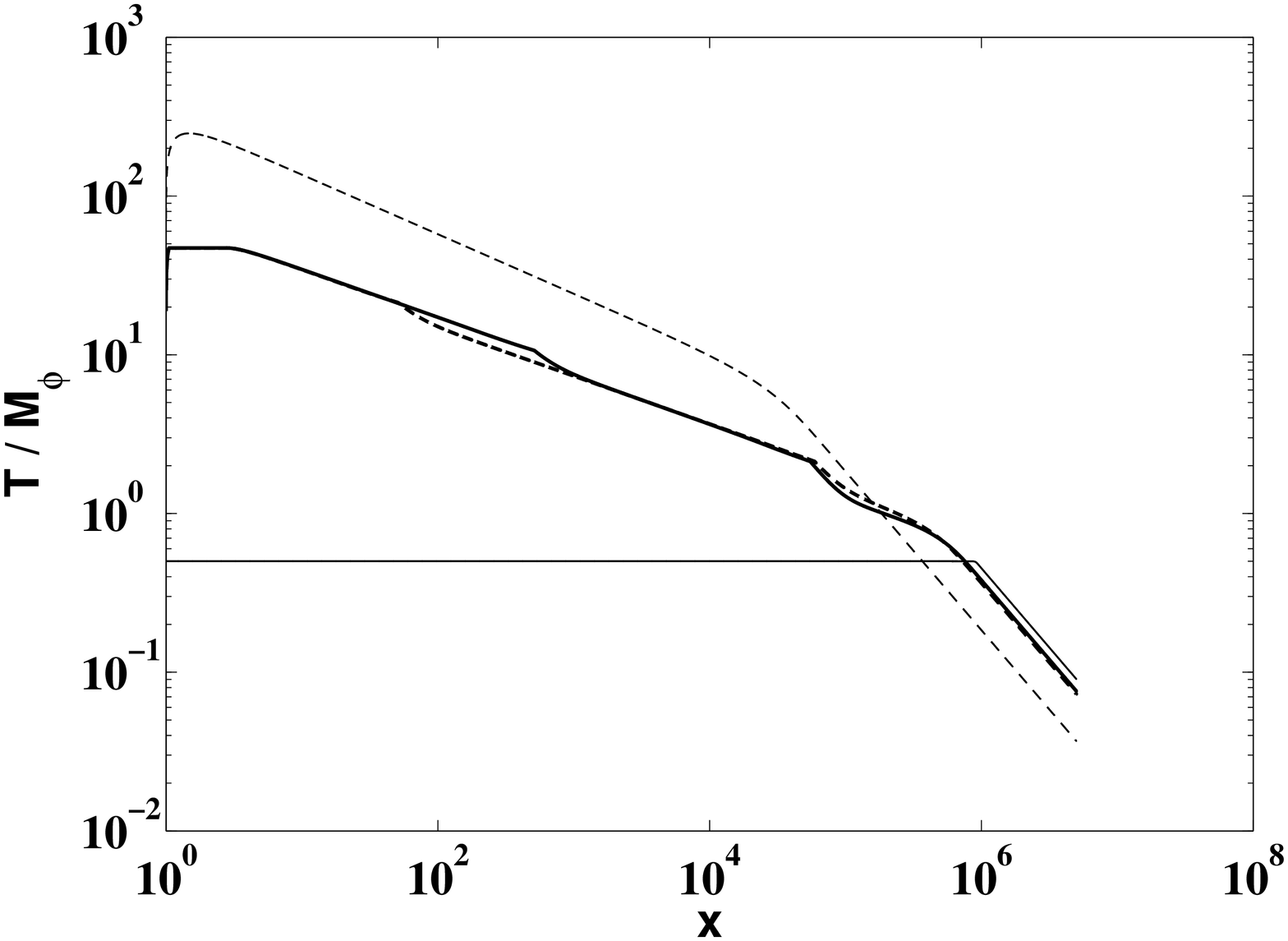}%
\end{center}

\caption[a]{Solutions for Boltzmann equations with two different sets of
  initial values: $g_{Y}^{2}=6\times 10^{-13}$, $M_\phi=10^{9}$GeV,
  $V^{1/4}=10^{14}$GeV (left) and $g_{Y}^{2}=3\times 10^{-9}$, $M_\phi=2\times
  10^{7}$GeV, $V^{1/4}=8\times 10^{11}$GeV (right). The curves are: results
  obtained by Kolb et al. in \cite{Kolb:2003ke} (thin solid); the case where
  thermal corrections are neglected, i.e. $\Gamma_\phi=g_Y^2M_\phi$ (thin
  dashed); the case where thermal corrections are included with $g_1=1$ and
  $g_2=0.1$ (thick dashed); and the case where thermal corrections are
  included with $g_1=1$ and $g_2=0.2$ (thick solid).}

\la{fig:boltz}
\end{figure}
The temperature of relativistic gas in an expanding universe is given by
\begin{equation}
\frac{T(x)}{M_{\phi}} = \left( \frac{30}{g_{*} \pi^{2}} \right)^{1/4}
\frac{R^{1/4}}{x}~,
\end{equation} 
where $g_{*} \approx 100$ counts the effective degrees of freedom. The time
evolution of temperature as a function of the scale parameter $x$, computed
with $\Gamma_\phi(T)$ of \eq{gammat} is plotted in \fig\ref{fig:boltz}
together with the results presented in \cite{Kolb:2003ke}. We have calculated
the temperatures for two sets of initial values following the cases II and III
discussed in \cite{Kolb:2003ke}, first with the scalar mass falling between
the reheat temperature and the maximum temperature reached by the universe
during the reheating, $T_{RH} < M_\phi < T_{MAX}$, with the reasonable
numerical values $g_{Y}^{2}=6\times 10^{-13}$, $M_\phi=10^{9}$GeV,
$V^{1/4}=10^{14}$GeV, where $V$ is the value of scalar potential at the end of
inflation; when the scalar mass is smaller than the reheat temperature,
$M_\phi < T_{RH}$, we adopt the values $g_{Y}^{2}=3\times 10^{-9}$,
$M_\phi=2\times 10^{7}$GeV, and $V^{1/4}=8\times 10^{11}$GeV.

As can be seen in \fig\ref{fig:boltz}, an era of a constant temperature which
was found by a partial consideration of the thermal effects relevant for the
decay of the inflaton condensate \cite{Kolb:2003ke}, actually vanishes when
one takes into account all the interactions between the scalar and the thermal
bath at the one-loop level. The inclusion of the absorption decay channel and
the possibility to decay into more than one type of fermion allow for the
reactions to continue at temperatures greater than the inflaton rest mass.
This effect is due to the existence of several fermionic species coupling to
the inflaton; in the case of one species only, the result would be almost
identical to the case of Kolb et al. \cite{Kolb:2003ke}, with only minor
changes due to the hole excitations.  As a consequence the thermal history of
the heat bath turns out to be qualitatively similar to the case when the
thermal masses are ignored. At the end of inflation the rapid decays of
inflaton raise the temperature to some maximum temperature, after which most
of the time the temperature scales as $T \propto a^{-3/8}$; however, there are
short periods during which the temperature drops faster because of absorption
channels reaching the level of fastest direct decay rates before closing with
decreasing temperature. If the thermal bath contained also bosons that couple
to the inflaton, the thermalization rate would be modified accordingly. The
details would again depend on the coupling strengths. Clearly, the
thermalization rate and the temperature of the resulting heat bath very much
depend not only on the details of the inflaton model itself but also on the
reaction rates of the decay products, which determine their plasma
masses. Such considerations might also be important for e.g. thermal
leptogenesis \cite{Fukugita:1986hr}.

%
\section*{Acknowledgements}

This work is partially supported by the Academy of Finland grants 75065 and
205800 and by the Magnus Ehrnrooth foundation. J.H. would like to thank
A. Jokinen for helpful discussions.


\end{document}